\documentclass[10pt]{iopart}
\usepackage{epsfig}
\begin{document}

\title{Overview of the Status and Strangeness Capabilities of STAR}
\author{P.G. Jones$^1$ and P.M. Jacobs$^2$ for the STAR Collaboration}
\address{$^1$University of Birmingham, UK.}
\address{$^2$Lawrence Berkeley National Laboratory, USA.}

\begin{abstract}
STAR is a large acceptance spectrometer capable of precision
measurements of a wide variety of strange particles. We discuss the STAR
detector, its configuration during the first two years of RHIC
operation, and its initial performance for Au+Au collisions. The
expected performance for strangeness physics and initial
data on strange particle reconstruction in Au+Au collisions are
presented.
\end{abstract}

\section{The Status of the STAR Experiment}

\begin{figure}[t]
\centering\epsfig{figure=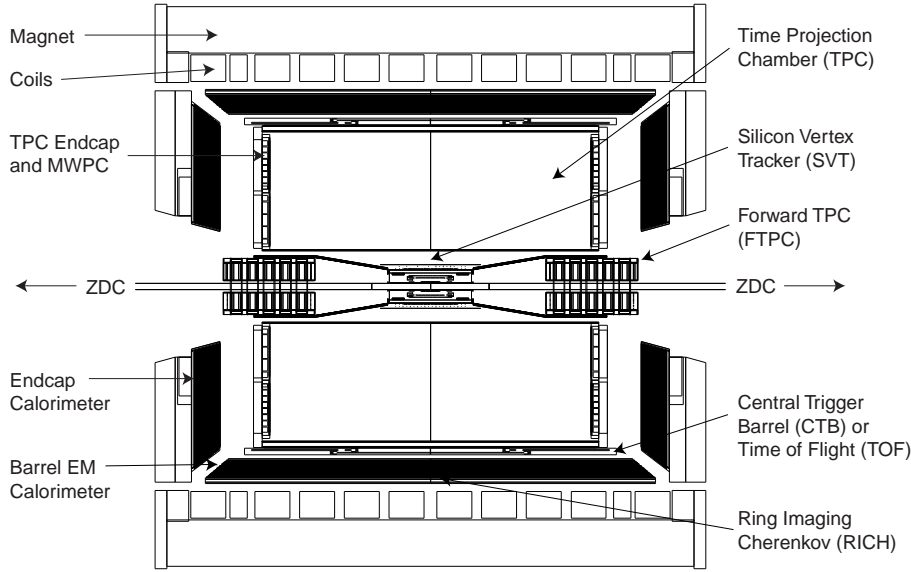, width=12cm}
\caption{Side view of the STAR detector configuration in the first
         two years of RHIC operation.}
\label{setup}
\end{figure}

The STAR detector at RHIC~\cite{STAR} is a large acceptance
spectrometer with a very broad physics program, including precision
measurements of a wide variety of strange particles. The STAR
experimental configuration during the early years of RHIC operation is
shown in Figure \ref{setup}. For the first year of RHIC running, the
active detectors are the Time Projection Chamber (TPC), the Zero
Degree Calorimeters (ZDC)~\cite{ZDC}, the Central Trigger Barrel
(CTB), and the Ring Imaging Cerenkov Counter (RICH). The solenoidal
magnet has a maximum field strength of 0.5 Tesla but for the first
year is run at 0.25 Tesla. For operation in the second year, STAR will
install the Silicon Vertex Tracker (SVT), two forward TPCs (FTPC), a
time-of-flight (TOF) patch, and 26 modules of the Barrel
Electromagnetic Calorimeter (EMC), comprising 22\% of the full barrel
EMC which will be completed by 2003. Details on the acceptance of the
various detector components relevant for strange particle
reconstruction are given in the following section.

\begin{figure}[t]
\epsfig{figure=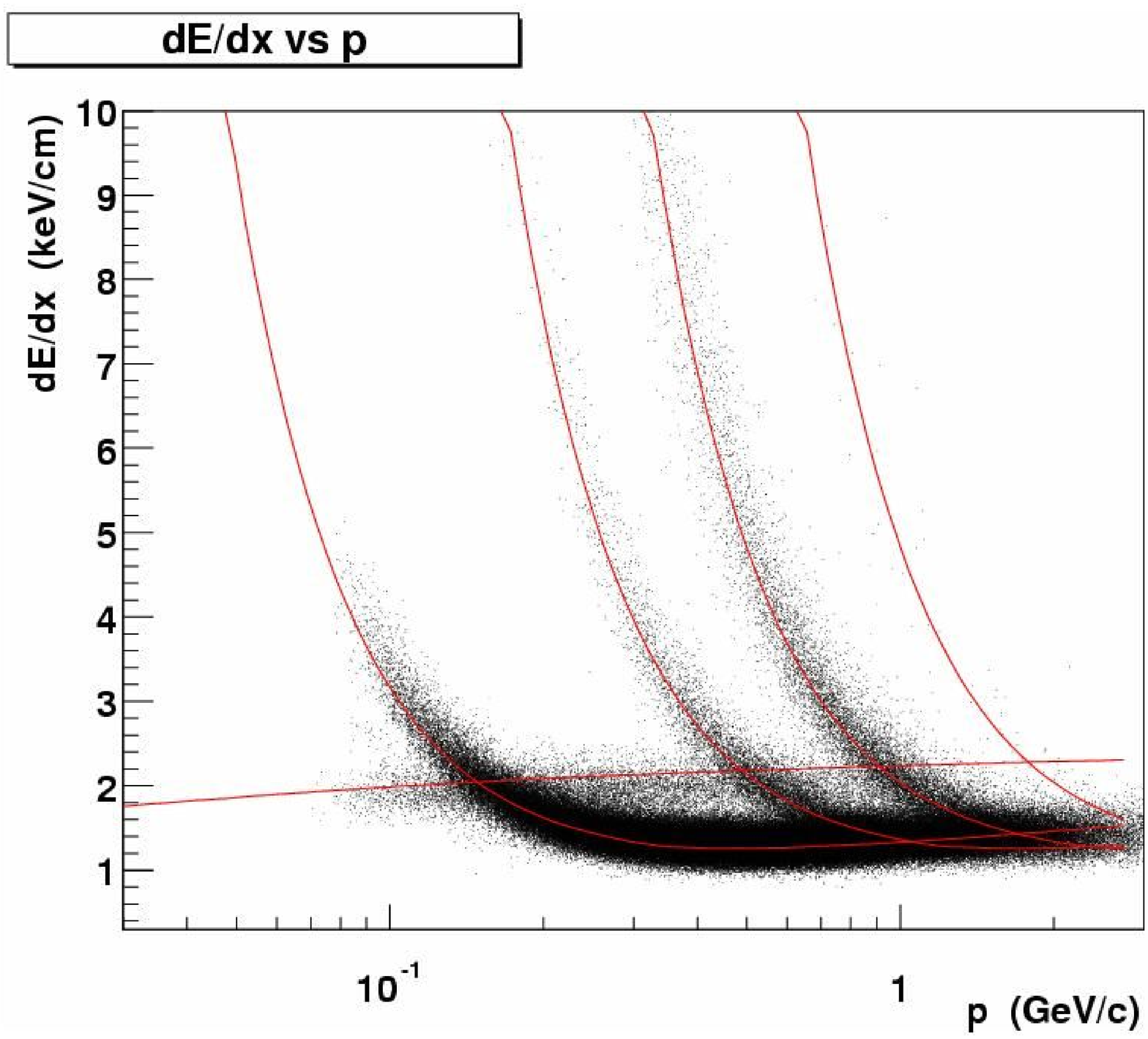, width=7cm}
\epsfig{figure=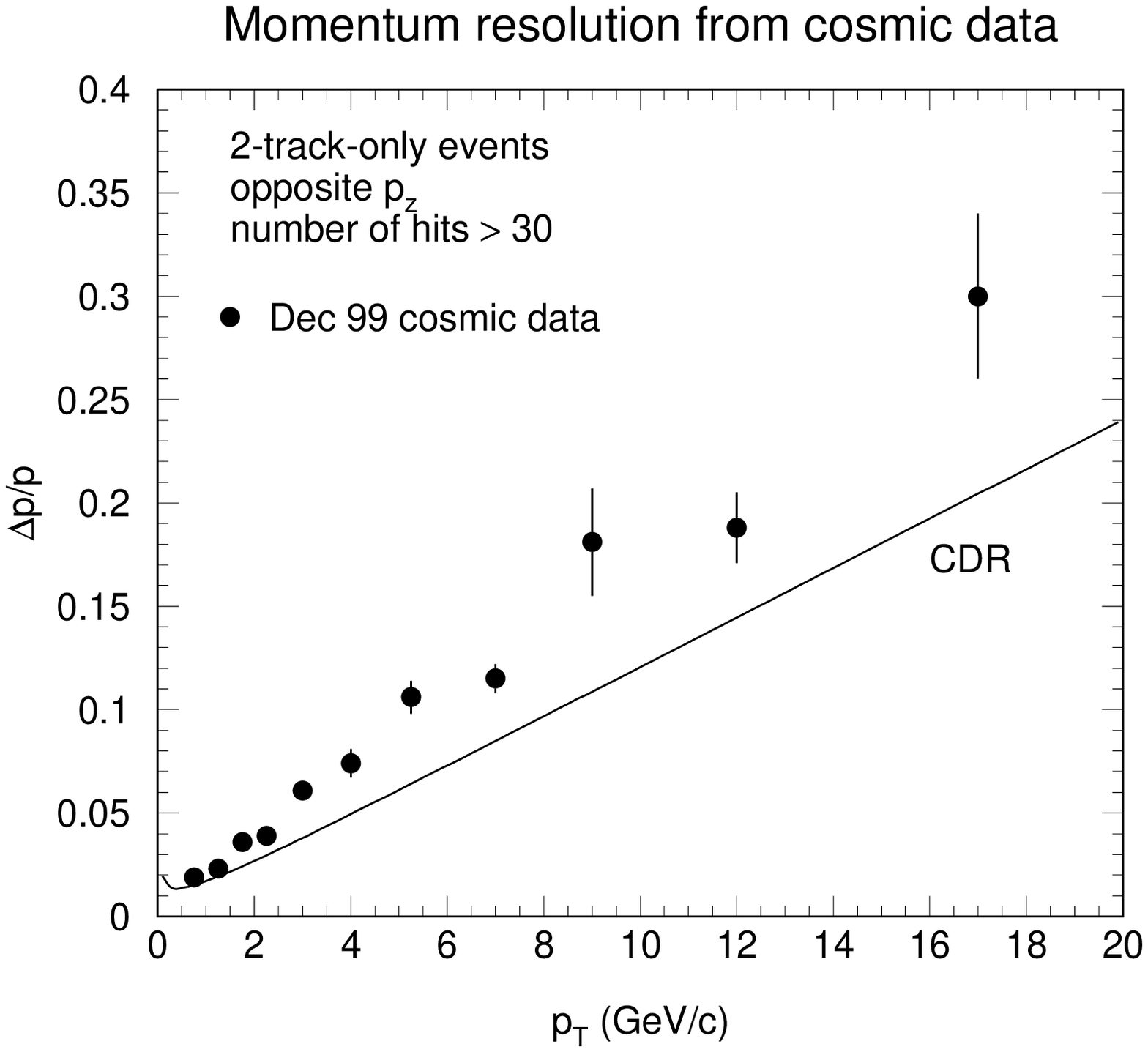, width=6cm}
\caption{
Preliminary data on TPC performance. Left panel: specific ionization
(dE/dx) versus track momentum for Au+Au collisions. Right panel: momentum
resolution in percent vs. momentum for cosmic rays reconstructed independently
in the upper and lower halves of the TPC. See text for further details.}
\label{TPCPerformance}
\end{figure}

RHIC began operations for physics in the summer of 2000. Together with
the other RHIC experiments, STAR observed collisions immediately.  The
quality of track reconstruction in the TPC is high and efforts to
understand and quantify the tracking efficiency are underway. The
performance of the STAR TPC is illustrated in
Figure~\ref{TPCPerformance}. The left panel shows the specific
ionization (truncated mean dE/dx) in the TPC for Au+Au collisions at
$\sqrt{s_{NN}} = 130$ GeV, as a function of track momentum. The curves
are the Bethe-Bloch parameterization of dE/dx. The bands for pions,
kaons, protons, deuterons and electrons are visible. The relative
dE/dx resolution is approaching the design goal of 8\%. The left panel
shows the momentum resolution of the TPC as a function of momentum for
charged tracks from cosmic rays that have been independently
reconstructed in the upper and lower halves of the TPC in a 0.5 Tesla
magnetic field. The solid curve labelled ``CDR'' is the design goal,
with the difference from measurement due in large part to known
instrumental effects that have since been rectified.

\begin{figure}[t]
\centering\epsfig{figure=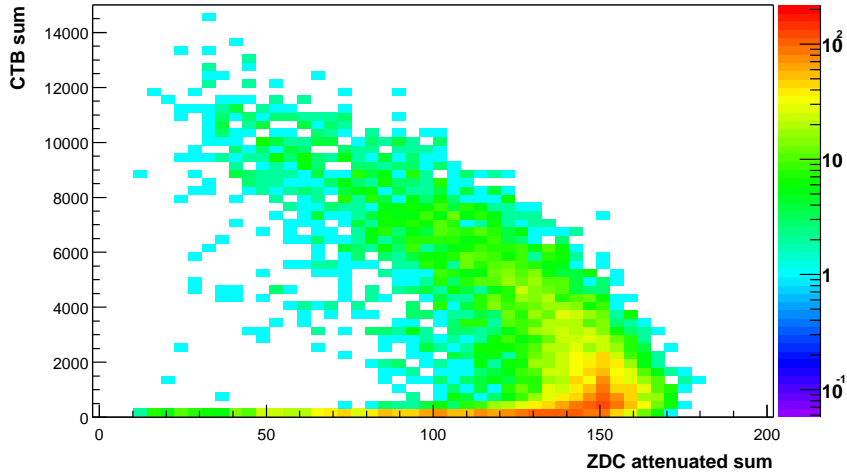, height=7cm}
\caption{Preliminary data from STAR Trigger detectors, showing total pulse
height in the CTB vs. summed pulse height in the two ZDCs.}
\label{STARTrigger}
\end{figure}

The response of the STAR trigger detectors (ZDC and CTB) to Au+Au
collisions at $\sqrt{s_{NN}} = 130$ GeV is shown in
Figure~\ref{STARTrigger}, under the condition that both ZDC signals are
above a threshold.\footnote{The threshold is set to eliminate
coincidences due to electronic noise but not those due to single
neutrons at beam energy.} The vertical axis is the summed pulse height
of all CTB slats and horizontal axis is the summed pulse height of the
two ZDCs. The ZDC measures forward-going neutrons ($\theta<2$ mrad)
and its response contains contributions from both nuclear collisions
and mutual Coulomb dissociation~\cite{BaltzEtAl} (the latter process
does not generate tracks at mid-rapidity and is not visible in the
figure). For nuclear collisions, low ZDC response corresponds to a low
number of spectator neutrons, which can occur for both very peripheral
and very central collisions, whereas the CTB signal increases
monotonically with increasing multiplicity at mid-rapidity. Thus, the
most central nuclear collisions correspond to low ZDC and high CTB
response (corresponding to high multiplicity at mid-rapidity).
The trigger inefficiency
for central collisions due to the coincidence condition applied to the
ZDCs was measured to be much less than 1\%.

The RICH is a proximity-focusing ring imaging Cerenkov counter for
identifying charged hadrons at high momentum within an acceptance of
10 msr at y$_{CM}$=0, and was developed as a joint project with the
ALICE collaboration. Rings with the expected photon yields have been
observed in the RICH for Au+Au collisions.

In summary, the STAR detector for the first year of RHIC operation is
performing extremely well, and we expect many physics results from
this first round of data taking.

\section{Strange particle measurements accessible to STAR}

By virtue of its large tracking acceptance and the ability to perform particle
identification over a wide kinematic range, STAR is well suited to measuring a
variety of strange particles. Table \ref{parts} lists the strange particles and
resonances that are accessible to STAR and which have been studied in 
Monte-Carlo simulations. Based on the result of these simulations we discuss 
the expected reconstruction performance of the STAR detector for various
strange particle measurements. 

\begin{table}
\begin{center}
\begin{tabular}{|l|c|c|c|} \hline
Particle        & Decay Mode           & Branching & Lifetime  \\ 
                &                      &   (\%)    & (cm/$c$)  \\ \hline\hline
$K^{\pm}$       & $\mu^{\pm}\nu_{\mu}$ &  64       & 371       \\
$K^{0}_{s}$     & $\pi^{+}\pi{-}$      &  68       & 2.67      \\
$\Lambda$       & $p\pi{-}$            &  64       & 7.89      \\
$\Xi^{-}$       & $\Lambda\pi^{-}$     & 100       & 4.91      \\
$\Omega^{-}$    & $\Lambda K^{-}$      &  68       & 2.46      \\ \hline
$\phi$          & $K^{+}K^{-}$         &  64       & Resonance \\
$K^{0*}$        & $K^{\mp}\pi^{\pm}$   &  67       & Resonance \\ \hline
\end{tabular}
\end{center}
\caption{List of some of the strange particles and their decay modes 
         accessible to STAR.}
\label{parts}
\end{table}

\subsection{Charged kaon reconstruction}

About 70\% of the strange quarks produced in a heavy-ion collision
at RHIC are carried by the kaons. Charged kaons can be identified in
STAR using a number of different techniques. The measurement of the
specific ionization of tracks in the TPC (dE/dx) gives an excellent
separation of kaons from protons and pions up to a momentum of around
800 MeV/$c$ (Figure ~\ref{TPCPerformance}). Charge kaons can also be
identified in the TPC via their decay (see Table \ref{parts}). In
principle, using this method it is possible to extend the measurement
of kaons out to a transverse momentum of around 5 GeV/$c$.


The identification of charged kaons at mid-rapidity in the range $1.1 <
p_{T} <3.0$ GeV/$c$ is performed by the RICH. In year two, the
charged kaon measurement will be augmented by the addition of a
time-of-flight patch, which will replace part of the central trigger
barrel and provide acceptance in the range $0.3 < p_{T} < 1.5$
GeV/$c$.

\subsection{Hyperon reconstruction}

Hyperons can be reconstructed in the central TPC from their charged
decay products. $\Lambda$ and $\overline{\Lambda}$ are found by
considering all pairs of oppositely charged tracks in the TPC which
are compatible with having originated from a vertex separated from the
primary interaction.  Neutral kaon decays, $K^{0}_{s}$, are also
reconstructed in this way.  Geometrical cuts are used to eliminate the
majority of the combinatorial background. The most effective cuts are
on the separation of the decay from the primary vertex and on the
impact parameter of the parent and daughters to the primary vertex. An
important cut for $\Lambda$ ($\overline{\Lambda}$) is on the specific
ionization information (dE/dx), which selects tracks compatible with being
protons (anti-protons).

\begin{figure}[t]
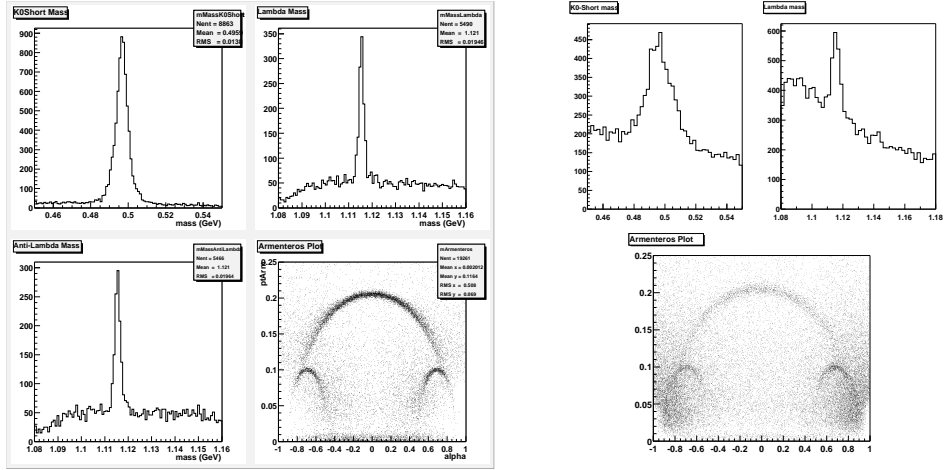

\centering
\begin{tabular}{p{.54\linewidth}p{.39\linewidth}}
{\vspace{-6cm}\epsfig{figure=masses_cuts_2923_hij.eps, height=6.5cm, angle=-90}} &
{\epsfig{figure=NewRealV0s.eps, height=6cm}}\\
\end{tabular}
\caption{The invariant mass distribution of $K^{0}_{s}$, $\Lambda$, and 
         $\overline{\Lambda}$ from 3000 HIJING events (left panel) and
         the preliminary $K^{0}_{s}$ and $\Lambda$ mass distributions
         from approximately 
         16000 minimum bias events of the current run (right panel).}
\label{v0}
\end{figure}

Figure \ref{v0} shows the invariant mass and Podolanski-Armenteros
plots~\cite{ArmPod} for 3000 HIJING central Au+Au events and approximately
16000 minimum bias events from the current run.  In both cases, no dE/dx
information has been used.  The analysis of the HIJING simulations found a
reconstructed yield of approximately 2 $K^{0}_{s}$ and 0.3 $\Lambda$
($\overline{\Lambda}$) per event. The same cuts, which were tuned to
optimise the signal to noise ratio in the Monte-Carlo, were applied to
the real data. A clear signal is observed. The background in the real
data is larger than expected from the simulation and this has since been
understood as due in part to an instrumental effect and an error in
the way one of the cuts was defined, both of which have been
corrected.

\begin{figure}[t]
\epsfig{figure=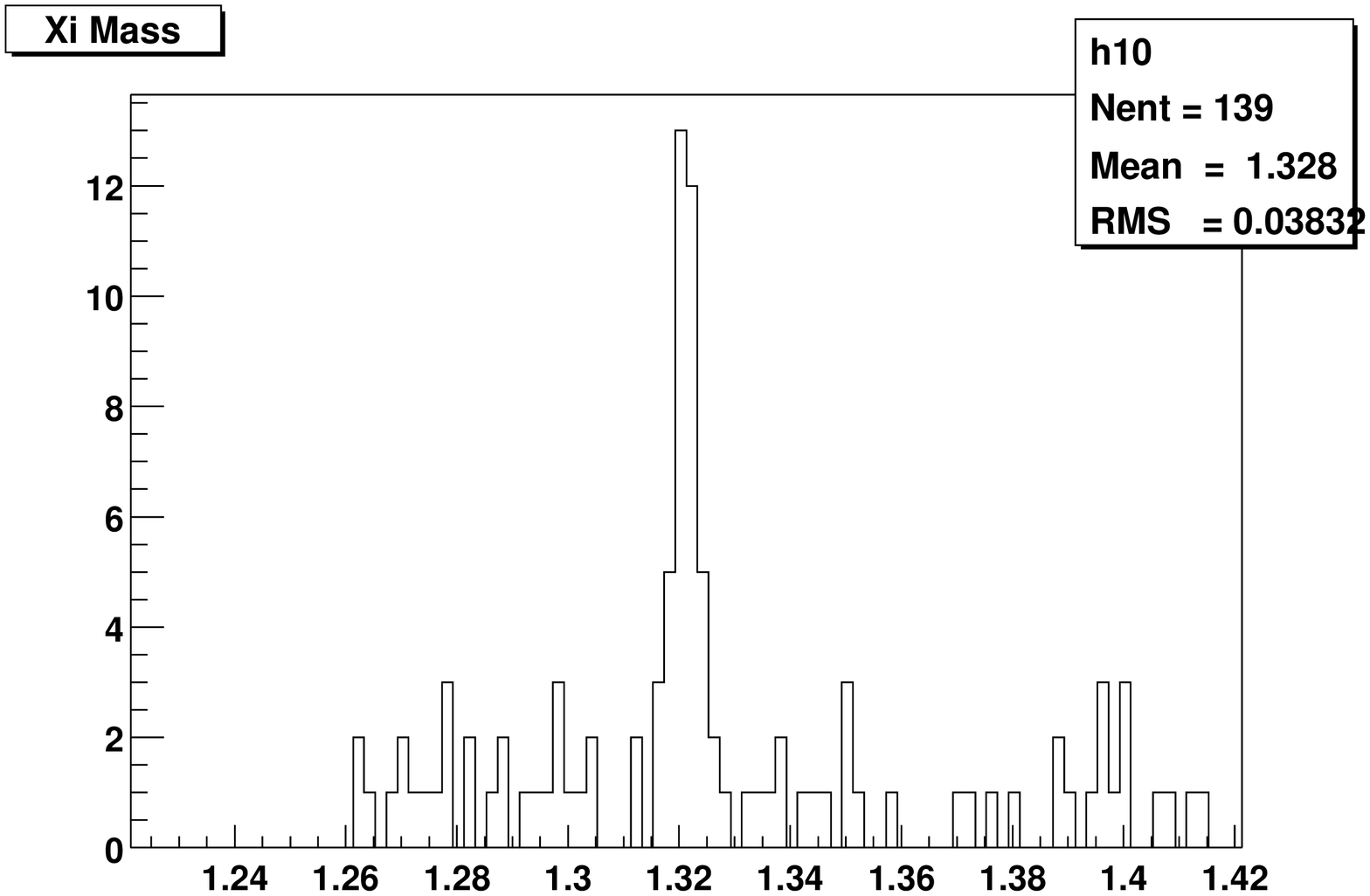, height=6cm, angle=-90}
\hspace{0.5cm}
\epsfig{figure=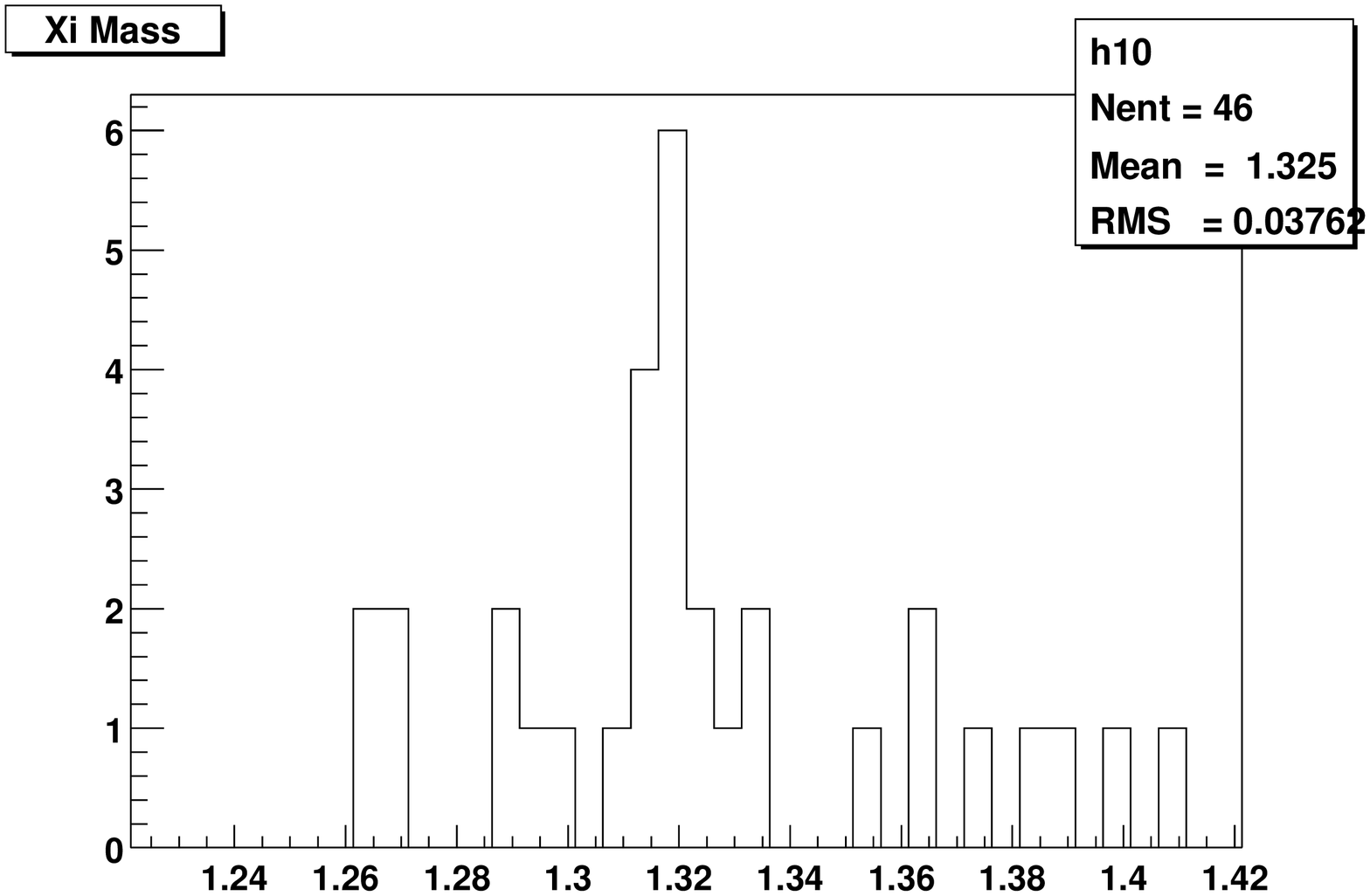, height=6cm, angle=-90}
\caption{The invariant mass distribution of $\Xi^{-}$ and $\overline{\Xi}^{+}$
         from 7000 HIJING central Au+Au events (left panel) and the
         preliminary distribution from approximately
         16000 minimum bias events of the current run (right panel).}
\label{xi}
\end{figure}

$\Xi$ reconstruction has also been studied in STAR. Signal
extraction in the year one configuration has been studied using events in
which an enhanced $\Xi$ signal had been added.
By limiting the acceptance to $\Xi$ with transverse momentum
$p_{T} > 0.8$ GeV/$c$, it was found that nearly all the combinatorial
background could be eliminated. The cuts used were similar to those
found useful for extracting $\Lambda$ and $K^{0}_{s}$.
Figure \ref{xi} shows the result of applying these cuts to 7000
reconstructed HIJING events. Also shown is a hint of a peak in the
experimental data obtained from 16000 minimum bias events from the
current experimental run. In both cases, the sum of $\Xi^{-}$ and
$\overline{\Xi}^{+}$ is shown.

In the second year of RHIC operation, the efficiency of hyperon 
reconstruction will be significantly enhanced by the addition of the 
Silicon Vertex Tracker. This detector comprises three layers of Silicon 
Drift Detectors at approximately 5, 10 and 15 cm radius from the 
nominal beam axis. (An additional layer of Silicon Strip Detectors at a
radius 25 cm will be added in year three). Due to the superior two-track 
resolution of the SVT, it should be 
possible to extend the measurement of hyperons to lower transverse 
momentum, where the combinatorial background for tracks reconstructed 
in the TPC alone becomes prohibitive. It will also make possible the 
measurement of the rare $\Omega$ baryon. 

The two Forward Time Projection Chambers (FTPC) will also be installed
for the second year of RHIC operation. These are novel radial drift
devices operating in the challenging tracking environment close to the
beam, inside the inner field cage of the main TPC. A feasibility study
has shown momentum resolution to be sufficient for $\Lambda$
reconstruction. This will add new acceptance in the
range $2.75 < y_{\Lambda} < 3.5$.

\subsection{Resonance reconstruction}

The reconstruction of strange particle resonances is also possible
in STAR. Resonance production presents an important test of thermal
production models and can be sensitive to the properties of the state
of matter in which they are produced. The $\phi$ meson is of
particular interest as a probe of chiral symmetry restoration. STAR
measures the $\phi$ via the charged kaon channel, $\phi \rightarrow
K^{+}K^{-}$.  This
measurement uses the TPC dE/dx information to select kaons and the
background is determined by mixing positive and negative tracks from
different events having approximately the same total multiplicity.
Other resonances have also been studied. A simulation of
5000 HIJING events, for example, has shown that it is should be possible
to reconstruct the $K^{0*}$(892) from year one data.


\section{Summary and Outlook}

The STAR experiment has begun taking data for physics and is
performing very well.  STAR will measure integrated yields and
particle ratios of strange particles in collisions ranging from p+p to
Au+Au. Table \ref{results} shows the expected reconstruction
performance based upon the HIJING model's prediction for the strange
particle yields in central Au+Au collisions.  Based upon these
results, a measurement of all but the rarest multiply strange baryons
should be possible in the first year of RHIC operation. Taking
advantage of the complete azimuthal coverage of the STAR TPC,
it might even be possible to perform
$K^{0}_{s}K^{0}_{s}$ interferometry. With the addition of the SVT in
year two, we will be able to measure the production of the $\Omega$
baryon and with increased efficiency at lower transverse momentum, it
may also be possible to study $\Lambda\Lambda$ correlations.

\begin{table}
\begin{center}
\begin{tabular}{|l|c|c|c|c|} \hline
Particle        & HIJING ($4\pi$)  & $> 9$ TPC hits & \multicolumn{2}{c|}{Reconstructed}\\
\cline{4-5}
                &                  &                & TPC    & TPC+SVT+SSD \\ \hline
$K^{0}_{s}$     & 250              &  30            & 2.0    & 6.5         \\
$\Lambda$       & 80               &  5             & 0.3    & 1.5         \\
$\Xi^{-}$       & 5                &  0.3           & 0.05   & 0.01-0.02   \\
$\Omega^{-}$    & 0.05             &  0.03          &        &             \\ \hline
\end{tabular}
\end{center}
\caption{The expected reconstructed yield of various strange particles
         per central Au+Au collision, obtained from the HIJING model.}
\label{results}
\end{table}


\end{document}